# A multidisciplinary framework for deconstructing bots' pluripotency in dualistic antagonism

Wentao Xu
Email: myrainbowandsky@gmail.com
Department of Science and Technology Communication,
University of Science and Technology of China, China.

Kazutoshi Sasahara
Email: sasahara.k.aa@m.titech.ac.jp
School of Environment and Society, Tokyo Tech, Japan.

Jianxun Chu
Email: chujx@ustc.edu.cn
Department of Science and Technology Communication,
University of Science and Technology of China, China.

Bin Wang
Email: ialsolikefishandchips@gmail.com
Independent researcher

Wenlu Fan
Email: wenlufangfang@gmail.com
Department of Science and Technology Communication,
University of Science and Technology of China, China.

Zhiwen Hu

Email: huzhiwen@zjgsu.edu.cn
School of Computer Science and Technology, Zhejiang Gongshang University, China.
Collaborative Innovation Centre of Computational Social Science, Zhejiang Gongshang University, China.

## Abstract

Anthropomorphic social bots are engineered to emulate human verbal communication and generate toxic or inflammatory content across social networking services (SNSs). Bot-disseminated misinformation could subtly yet profoundly reshape societal processes by complexly interweaving factors like repeated disinformation exposure, amplified political polarization, compromised indicators of democratic health, shifted perceptions of national identity, propagation of false social norms, and manipulation of collective memory over time. However, extrapolating bots' pluripotency across hybridized, multilingual, and heterogeneous media ecologies from isolated SNS analyses remains largely unknown, underscoring the need for a comprehensive framework to characterise bots' emergent risks to civic discourse. Here we propose an interdisciplinary framework to characterise bots' pluripotency, incorporating quantification of influence, network dynamics monitoring, and interlingual feature analysis. When applied to the geopolitical discourse around the Russo-Ukrainian conflict, results from interlanguage toxicity profiling and network analysis elucidated spatiotemporal trajectories of pro-Russian and pro-Ukrainian human and bots across hybrid SNSs. Weaponized bots predominantly inhabited X, while human primarily populated Reddit in the social media warfare. We analysed more than 20M tweets and 1M Reddit comments and found that English content on both platforms exhibited Granger causality with cascade size and depth in minor languages like Japanese and German ($p<0.05$). Temporal correlations between humans and bots posting frequencies heightened English toxicity over minor languages. This rigorous framework promises to elucidate interlingual homogeneity and heterogeneity in bots' pluripotent behaviours, revealing synergistic human-bot mechanisms underlying regimes of information manipulation, echo chamber formation, and collective memory manifestation in algorithmically structured societies.

## Introduction

In 1988, artificial intelligence (AI) pioneer Marvin Minsky proclaimed that "The question is not whether intelligent machines can have any emotions, but whether machines can be intelligent without any emotions"[1]. Approximately 15 years later, the major contemporary social networking services (SNSs), including Facebook, X (former Twitter), Reddit, LinkedIn, emerged. In recent years, as AI technology has advanced rapidly, humanoid automated accounts controlled by sophisticated algorithms, also known as social bots, have proliferated across various SNS platforms[2,3,4,5,6]. An important issue concerns comprehensively understanding the role of social bots in situations of dualistic antagonism on SNSs, where there are two opposing factions with staunchly conflicting views. Previous studies have examined political dynamics on X, with their findings sometimes extrapolated to senarios[2,7]. However, such extrapolation is likely based on the implicit assumptions that all SNS platforms constitute homogeneous ecosystems. In reality, integrating diverse multilingual data from multiple online communities across SNSs presents a unique analytical challenge that has not been fully addressed.

This study proposes a theoretical framework to examine dualistic antagonism on SNSs. We adapt the concept of "pluripotency"[8] from stem cell biology to characterise the diverse roles of social bots in online polarization. Pluripotency refers to the ability of embryonic stem cells to differentiate into multiple cell types. Similarly, social bots exhibit pluripotency through their capacity to carry out various tasks on SNS. Prior research has shown social bots disseminating fake news[2,9,10,11,12], retrieving private information of online users[13,14,15,16], influencing the attention dynamics of online users[17,18,19,20,21], engaging with human users to formulate their positions[3,22,23,24]. We conceptualize the pluripotency of social bots as their AI-driven morphological ability, encompassing three attributes: adaptation, intelligence, and distribution (AID). Adaptation indicates social bots can survive and gradually adjust to homogeneous SNS environments. Intelligence suggests bots' capabilities are evolving. For example, recent findings suggested that social bots, empowered by large language models (LLMs), have the capacity to deteriorate the popular bot-detection methods[25,26]. Distribution means bots can strategically position themselves in social networks[27] and interact effectively with humans[28,29,30]. This AID framework characterises bots' dynamic, anthropic online behaviours driven by informational and network signals. Their pluripotency allows bots to take on different forms and roles that may encourage or exacerbate antagonism on polarized issues. Understanding bots' AID attributes provides an important context for studying dualistic antagonism on contemporary SNSs.

While prior work has examined polarized antagonism emerging from homogeneous social dynamics, SNSs exhibit meaningful heterogeneity that warrants consideration. It is true that homophily, the tendency for individuals to connect with others similar to themselves, influences interactions among like-minded users like journalists in social networks[31]. However, computational studies of networks like BlogCatalog, Last.fm, and LiveJournal found homophily is not the dominant factor in forming new ties[32]. Platforms such as Reddit and X also display heterogeneity in their structures and users. Analyses show political Redditors exhibit diverse behaviours[33], and user retention on X correlates with contact diversity[34]. The well-being effects of social media differ across platforms and demographic groups, indicating impact heterogeneity[35]. Researchers' motivations and goals also vary by platform used[31]. Moreover, platforms balance multiple, sometimes conflicting values in how they construct data value, allowing for divergent interpretations and participatory modes[35]. Together, characterizing SNSs as homogeneous overlooks real functionality, impact, and value heterogeneities across these diverse platforms. A more nuanced perspective is needed to understand phenomena like dualistic antagonism, accounting for heterogeneity in interpretive frameworks across SNSs.

The heterogeneous nature of SNSs enables users to fulfil diverse niche interests by posting varied content across different platforms. The same individual may exhibit preferences for distinct posting and networking behaviours on disparate SNS. Consequently, social media has become integral to numerous disciplines beyond their initial communications functions. SNSs provide a multidisciplinary domain encompassing areas like politics, economics, linguistics and mass media. They serve as an online forum for political discourse where language conveys ideologies and social practices[36]. Furthermore, social media facilitates interdisciplinary collaboration through sharing data, information and knowledge via technology. This allows for addressing real-world issues from diverse vantage points.[37]

Overall, social media's capacity to foster transdisciplinary communication, interaction and exchange of concepts renders it a multidisciplinary space.

While many social media studies have focused on political events like 2016 and 2020 U.S. presidential elections on X[2,38,39,40,41], more research is needed to comprehensively understand social bots' diverse roles across competing online discussions. Studies of elections employ data analysis to elucidate campaign dynamics but represent only a narrow view. Important debates beyond politics have also emerged on issues like vaccination[42], BlackLivesMatter[43], and QAnon[44], and sports games as long as anti-homophily exists[45]. These topics resemble dualistic antagonism in that they inspire polarization yet diverge from homophily. Such conversations critically shape collective memory. Collective memory, as manifested through shared historical understandings, is integral to national identity formation. It communicates past information and experiences, establishing foundations for collective self-perception[46]. Evidence suggests collective memory may exacerbate belief in conspiracy theories through intergroup effects[47]. For example, "$n$G" (from 2G to 5G) conspiracy narratives appealing to technological anxieties have proliferated online for decades[48]. This indicates strong collective memory might fuel polarized online debates, as competing narratives vying for acceptance mobilize public opinion.

Spatiotemporal features play an important role in analysing information diffusion processes across social media platforms. The temporal ordering embedded within spreading sequences uncovers hidden attributes and dynamics of networks pivotal to precisely forecasting subsequent message recipients[49,50]. By integrating network structures and temporal signals, researchers have successfully boosted cascade prediction model accuracy[49,51]. Moreover, the temporal reticulation of retweets delivers useful insights regarding diffusion dynamics such as intervals between retweets and transmissions within the follower network[52]. Understanding such patterns facilitates identifying influential users and key transmission routes empowering applications like viral marketing and optimized news dissemination[53]. Furthermore, modelling the likelihood of user re-sharing behaviours longitudinally utilizing hazard functions has proven adept at anticipating cascade growth and simulating viral propagation[54].

Here, we applied our proposed framework to analyse information operations during the Russo-Ukrainian war (**Fig. 1**). This ongoing conflict represents a paradigm shift in modern social media warfare. Tensions have existed since 2014, escalating on 24 February 2022 with Russia's "special military operation" in Ukraine. A substantial portion of SNSs users discussing this topic are bots. They have been implicated in spreading misinformation, fake news, conspiracy theories and misleading narratives[55]. Early in the war, over 20% of pro-Russian accounts were found to be bots[56]. The cognitive dimension of conflict has expanded from traditional spheres like physical battlefields, television and the Internet to massive SNSs like X and Reddit. Within our framework, bots are characterised by their influencing capacity via cascade size and depth over time. Co-retweeting and co-reply networks reveal human-bot engagement dynamics within shorter windows. Linguistic features like toxicity levels further examine interactions and correlations between communities, validating the overall characterization. Bots' evolved pluripotency allows them to adapt across languages and platforms, moving from information dissemination to incorporating large language models. The unlawful killings in Bucha provide an empirical case study to demonstrate our framework's real-world applicability in decoding information operations.

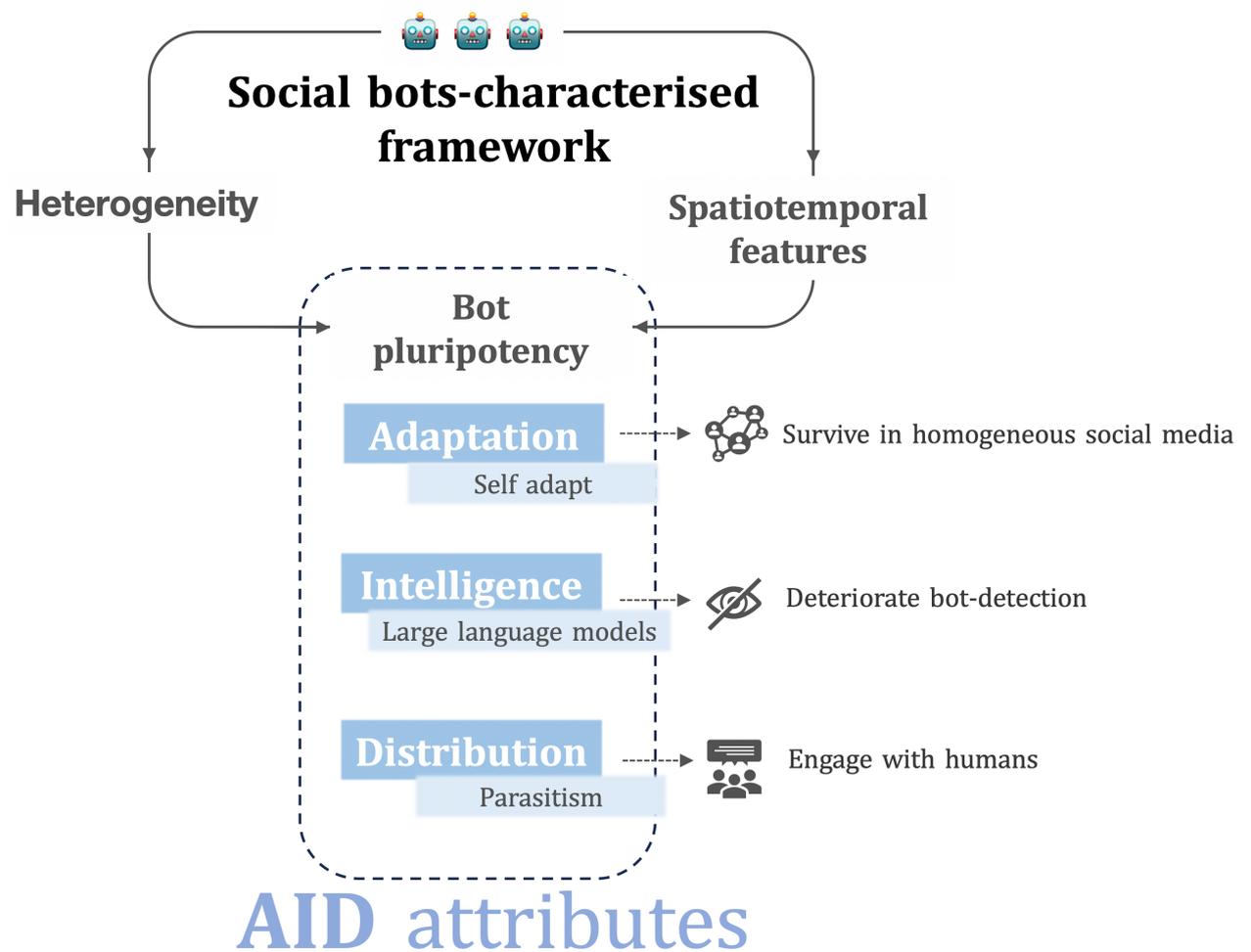

**Fig. 1 | Proposed framework for profiling social bots fuelling dualistic antagonism across social networking services (SNSs).** Both network heterogeneity and spatiotemporal dynamics collectively underlie bots' evolving "pluripotency". Adaptive capabilities allow automated accounts to self-adapt for homogeneous SNSs. Incorporating sophisticated natural language processing via large language models enhances bots' intelligence and undermining detection efforts. Strategic distribution patterns highlight bots' parasitic attributes through targeted engagement with human users. Analysing attributes of adaptation, intelligence and dissemination (AID model) provides insight into bots' multifaceted functions. Quantifying dimensions of the AID model across platforms over time offers a holistic perspective on bots' dynamics and impacts. The model considers technological, structural and behavioural factors

determining how bots exploit networked structures and human vulnerabilities to most effectively stoke antagonism.

Arguably, the patterns of human-bot engagement during the early stages of contemporary social media warfare remain not fully elucidated. Understanding these patterns is pivotal for decrypting misinformation propagation, information manipulation, polarization and echo chambers over the conflict's evolution. In addition to engagement, synchronized bot actions also merit investigation concerning amplifying misinformation, hate and polarization[57]. Examining bot actions amid the Russo-Ukrainian War could elucidate how bots synergistically operate to achieve social media warfare objectives. Linguistic features are important for evaluating human emotions and stances[58]. However, utilizing natural language processing techniques like LLMs for downstream tasks incurs high computational expenses. While previous work has analysed social bots under political contexts, developing a simple, lightweight empirical framework remains necessary to quantitatively profile bot pluripotency regarding contemporary wars across social media. Such a framework could facilitate a more comprehensive decoding of information confrontation in this novel cognitive battlespace.

# Results

## Polarized bots dominate Japanese and English language communities

The Botometer scores of the major users are equal or close to 0.8, indicating that the majority of the users are bots in both language communities (**Supplementary Fig. S1**). In addition, we found that the users of Japanese language community are significantly bot-like than their English counterparts (*t*-test, *p*=0.0)

We further obtain polarized the retweet network of each language community, and classified users as "pro-Ukraine bots", "pro-Ukraine human", "pro-Russia bots" and "pro-Russia human" (**Fig. 2**). In Japanese language community, there are 27,148 nodes and 194,690 edges, where there are 67.43% pro-Ukraine bots, 19.19% pro-Ukraine human, 10.95% pro-Russia bots and 2.42% pro-Russia human. In English language community, there are 100,084 nodes and 573,885 edges, where there are 57.72% pro-Ukraine bots, 31.58% pro-Ukraine human, 7.49% pro-Russia bots, and 3.21% pro-Russia human.

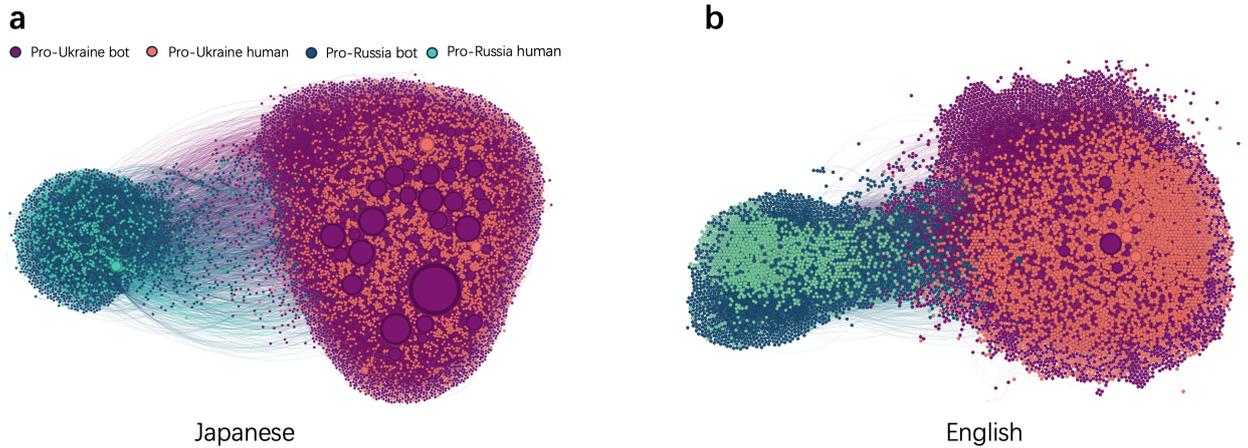

**Fig. 2 | Retweet networks (*3*-core) of Japanese (a), and English language community (b) on X.** The users are classified as pro-Ukraine bot, pro-Ukraine human, pro-Russia bot, and pro-Russia human in each community. The majority are pro-Ukraine bots in both languages. Note that the size of a node is scaled by its in-degree. The scalability is validated only in each RT network, respectively.

In the Japanese language community, the top three indegree users are "@nippon_ukuraina" (k=9,391) (a Ukrainian influencer living in Japan), "@KadotaRyusho" (a local writer & journalist),"@iii123123iii" (a film maker), while the English counterparts are "@KyivIndependent" (k=38,185) (a local Ukrainian online newspaper), "@DmytroKuleba" (k=18,708)" (Minister of Foreign Affairs of Ukraine), and "@olgatokariuk" (k=18,690) (an independent journalist and disinformation researcher). All of these top-indegree bots happen to be pro-Ukraine bots. These results suggest that the atmosphere of the social media war could be pro-Ukraine.

The retweeting behaviour is a crucial approach for bots to disseminate information. The pro-Ukraine bots tend to cluster with pro-Ukraine human on one side, while pro-Russia bots and pro-Russia human form clusters on the opposite side, which indicates that human and bots on the same side are highly correlated.

## Japanese bots are more influencing

When we analysed the retweet dynamics of four classes of users, including "pro-Ukraine human," "pro-Ukraine bot," "pro-Russia human" and "pro-Russia bot" of English and Japanese community. We found that bots diffused significantly more frequently, and more broadly than human in both language communities, and the user size of bots is much more than human.

The cascade size of Japanese and English (**Fig. 3**) shows a similar distribution in each corresponding user class. (Results of the Kolmogorov-Smirnov (K-S) test are shown in **Supplementary Table S1**). We further found that Japanese cascade size of pro-Ukraine bots is significantly greater than pro-Ukraine human, and the cascade size of pro-Russia bots is significantly greater than pro-Russia human as well. In addition, the cascade size of pro-Ukraine bots is significantly larger than pro-Russia bots in

Japanese language community. By contrast, the cascade size of English pro-Ukraine bots is significantly greater than pro-Ukraine human, but we did not find a similar tendency between pro-Russia bots and pro-Russia human in English. When we compared the interlanguage difference, the cascade size of Japanese pro-Ukraine bot was significantly greater than its English counterpart, and the cascade size of Japanese pro-Russia bot was significantly greater than English pro-Russia bot as well (Results of *t*-test are shown in **Supplementary Table S2**.)

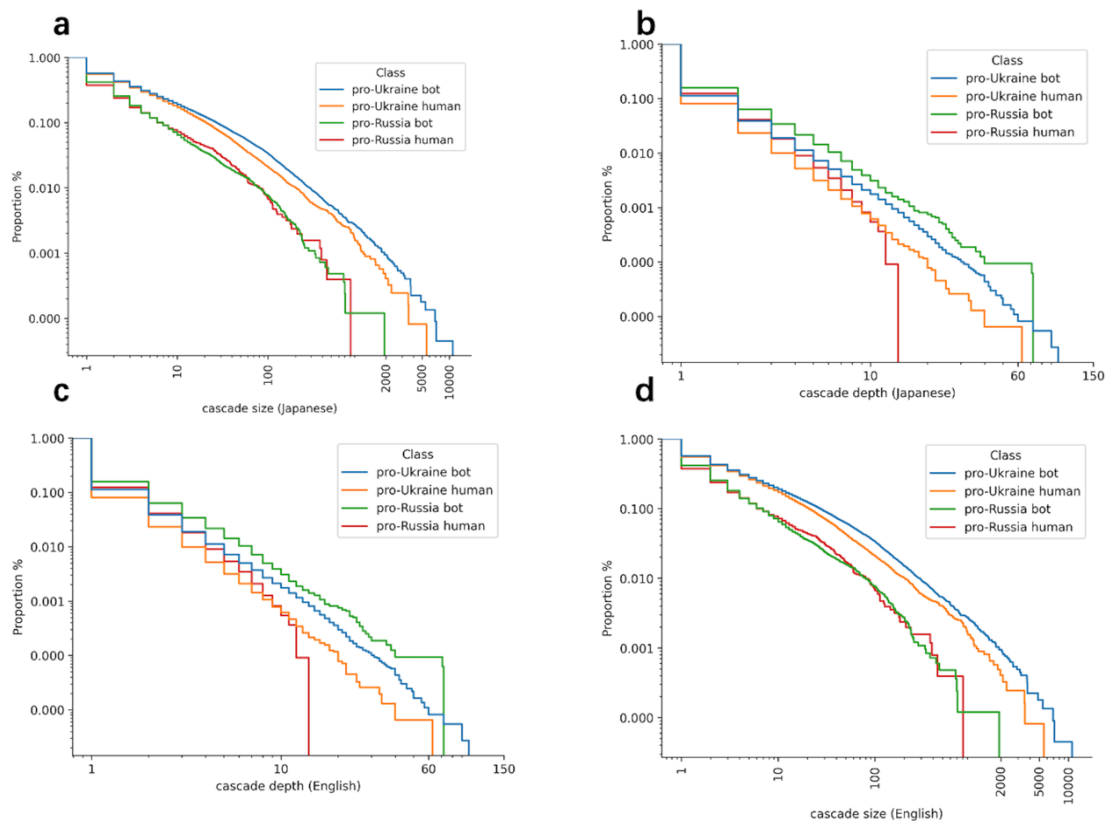

**Fig. 3 | Complementary cumulative distribution functions (CCDF) of cascade size and cascade depth of Japanese and English language communities. a-b**, Japanese language community, **c-d**, English counterparts.

In terms of cascade depth, pro-Ukraine bots retweeted significantly much more than pro-Ukraine human, and pro-Russia bots retweeted significantly much more than pro-Russia human in both languages (t-test, $p<0.001$). When we compared the interlanguage difference of #retweeting times, Japanese pro-Ukraine bots retweeted significantly more than their English counterpart, and Japanese pro-Russia bots retweeted significantly more than their English counterpart as well (t-test, $p<0.001$), as well. We also asked whether the cascade depth of English language community would Granger cause the counterpart of Japanese language community. Through the statistical results we rejected the null hypothesis and concluded that the English counts Granger causes Japanese counterpart, since all of the *p*-values are below the significance level of 0.05, and there is no reverse causality (**Supplementary Table S3**). It is reported that social bots accounted for a large portion of influential nodes in online conversations during major global events. Smart and his colleagues analysed the impact of bot activity on online discourse during the Russo-Ukrainian conflict and found that bots significantly influenced human[56]. Based on these findings, we assume that bots' behaviours of one side correlate with human of the same side, rather than the other side, because bots are likely to target influential human, increasing their exposure to negative information, and exacerbating social conflict online[59]. To verify this assumption, temporal patterns of retweet behaviours of human and bots were quantified.

The unique-user frequencies for each user class on the unique URL level illustrate that the patterns of user number increase follow similar trends across the time series (**Supplementary Fig. S2**). To statistically describe this observation, we measured the Pearson correlation coefficient of temporal oscillations of unique users retweeting unique URLs (**Supplementary Table S4**). Our results reveal that bots correlated with humans of the same side to a much higher degree than human of the other side of both languages.

While it is true that the English language community is considerably larger than the Japanese community, it would be overly simplistic to assert that the influence of the English language community is significantly greater than that of the Japanese community. Japanese social bots are more influential than English social bots in terms of the cascade size and the cascade depth, suggesting that X community of Japanese language community might be dominated by social bots. Next, we will investigate the penetrativeness between social bots and human.

## Japanese bots are more penetrative

To further investigate the bot-biased behaviour, we quantified the bot strategy of information spreading and examined the degree to which human reply upon and engage with the content produced by social bots. Three metrics including retweet pervasiveness (RTP), reply rate (RR), and human to bot rate (H2BR) were leveraged to measure the engagement. We evaluate each warfare side separately; thus, we compare the engagement between bots and human with the same side of different language communities. We described the results for each group of bots (**Supplementary Table S5**). Several diverse aspects merit consideration. We can observe that pro-Ukraine bots of Japanese language community are significantly more effective than their English counterparts. We found that the gap between the Japanese and English groups, with respect to all the metrics, is significant.

## Japanese bots are global players, but English bots are local

Co-retweeting, as mentioned, belonging to an umbrella of co-strategy, is a co-occurred retweeting behaviour by users within a short period of time. Here, the time window is set within 15~55 seconds, with an interval of 10 seconds, and the co-retweet networks are visualized in **Fig. 4**. We found that pro-Ukraine bots dominate both Japanese and English co-retweet networks across the five-time windows (**Supplementary Table S6**) and that the number of pro-Ukraine bots of English language community is significantly larger than Japanese community in each time interval. Different from the co-retweet network, co-reply network shows a human-bots "competing" network, where human tend to reply internally, and bots likewise (**Supplementary Fig. S3**). The number of pro-Ukraine rocketed to 2,850 in 15s time-window for the English language community, suggesting bots were connecting and integrated into a complex network for spreading viral messages. In contrast, the Japanese bots were much less connected at the beginning. However, following the 25s time window, Japanese bots became considerably more connected, but English bots were well-connected across time windows.

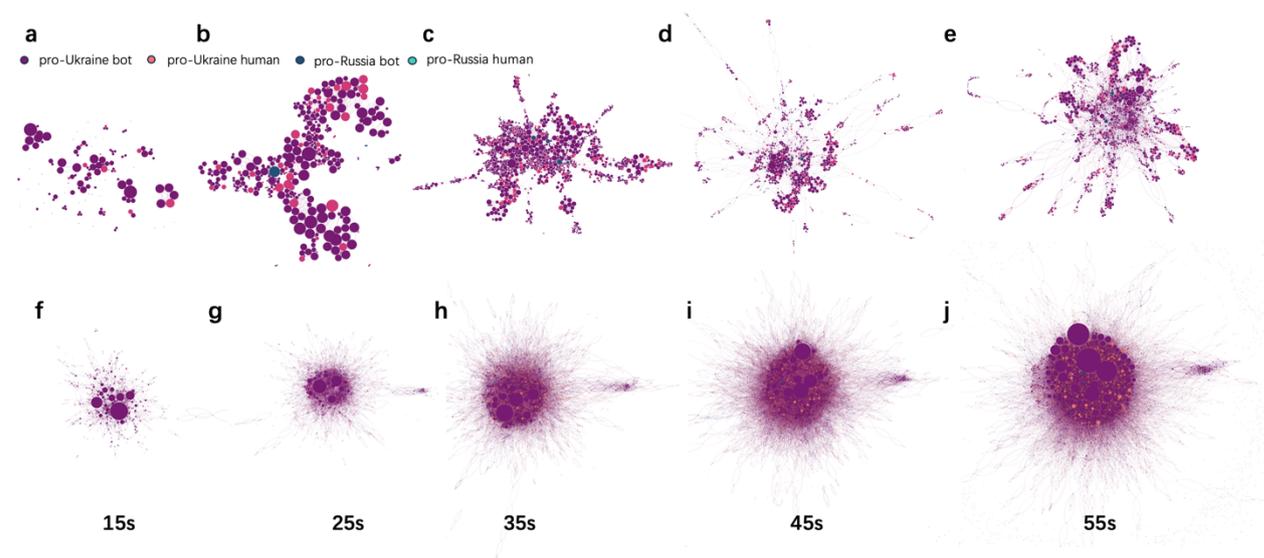

**Fig. 4 | Co-retweet network of Japanese and English language community.** The time windows range from 15s to 55s with an interval of 10s. Most of the co-retweeting users are bots in both language communities. Japanese language community started from single and small scatters to a more compact and larger cluster (**a-e**), whereas the English community evolved from a smaller cluster to a larger cluster (**f-j**). Note that the size of a node is scaled by its degree. The visualization scalability is validated only in each time window, respectively.

We then assumed that the importance of connected bots of the two language communities could be different. To investigate this assumption, we quantified the clustering coefficient against the centrality (degree) of each user (**Fig. 5**). Since the 35-second time window, it becomes apparent that a majority of users in the Japanese language community have maintained a consistent level of centrality when the clustering coefficient increased. The rising clustering coefficient as time goes on may reveal the users' tendency to interact more closely with a specific subset of other users, potentially forming smaller, highly connected communities within the Japanese language community. This could be due to the fast formation of the bots' network spreading of pro-Ukraine information. The stability in centrality among the majority of users suggests that the users continue to play key roles in maintaining the overall structure (**Fig. 5c-e**) and the connectivity of the community, contributing to the Japanese community's resilience of pro-Ukraine. In contrast, the centrality demonstrates an inverse relationship with the clustering coefficient in the English language community, suggesting that English users tend to promote local cohesion and connectivity within the community, but do not play a significant role in the global structure of the network. English co-retweet network formation is a transition process, during which, the pro-Ukraine bots dominated network structure with stronger local clustering shifts towards a more centralized and potentially polarized network structure, where certain pro-Ukraine bots are gaining influence, but the overall community structure is weakening.

Further statistical analysis showed that the degree centrality of Japanese is significantly larger than English (*t*-test, $p<0.001$), suggesting that the overall structure of Japanese co-retweet network is much more stable than English. This special structure may also explain why Japanese bots are more penetrative than English (**Supplementary Table S5**).

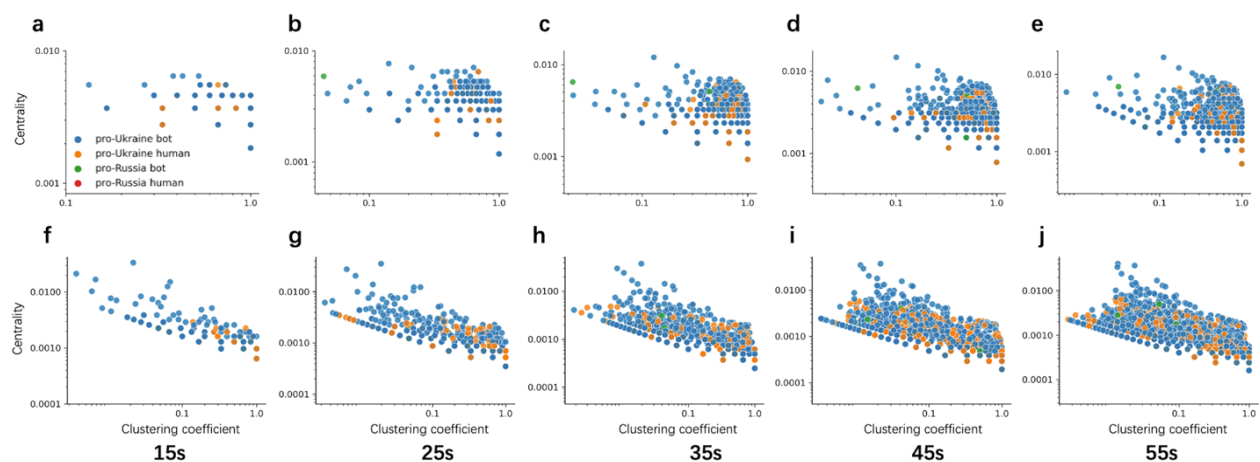

**Fig. 5 | Clustering coefficient against centrality (degree) of co-retweet network.** Centrality (degree) within the time window between 15s and 55s, with an interval of 10s. **a-e**, Japanese language communities, and **f-j**, English community.

Japanese pro-Russia users are more aggressive

To characterise the nature of bot linguistic capability, we measured the toxicity of languages used in the X text object and tested the differences of intra- and interlanguage community, across pro-Ukraine bot, pro-Ukraine human, pro-Russia bot and pro-Russia bot (**Fig. 6**). We used *t*-test to estimate the language toxicity difference.

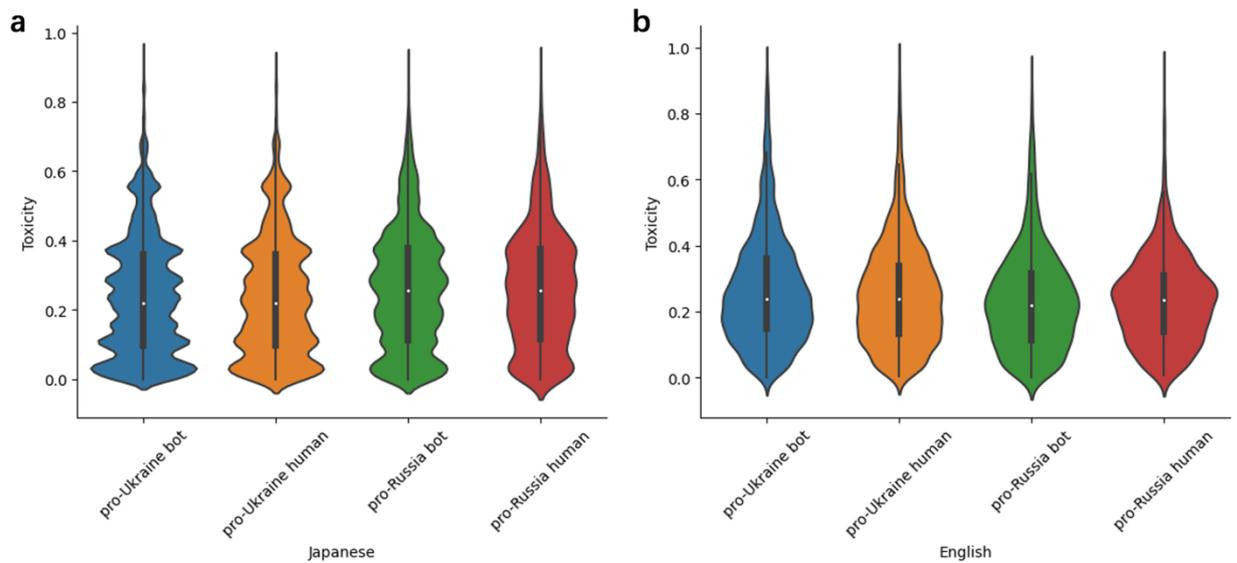

**Fig. 6 | Violin plots represent language toxicity of pro-Ukraine bots, pro-Ukraine human, pro-Russia bots and pro-Russia human of Japanese (a) and English (b) language community.**

Interlanguage analysis indicates that the overall toxicity of English text is significantly greater than that of Japanese (**Supplementary Table S7**). On a user-class pairwise level, the same significance is identified as well. Recall that Japanese bots obtained much more retweet pervasiveness, reply rate and human to bot rate (**Supplementary Table S5**), and that Japanese bots transit to a more stable, compact, and small connected community **(Fig. 4, Fig. 5)**. In such a low-openness ecosystem, once toxic contents pollute community, the polluted contents might help develop an echo chamber about the social media war. Remember that bots are more penetrative than human. The high penetrativeness of bots would attract more human audiences. We thus, assume that human probably followed bots on the same side rather than the other side. For example, the intention of pro-Ukraine bots is assumed to spread Ukraine-supported information and attract more users to support Ukraine, during the information warfare. To examine this assumption statistically, we compare the text toxicity between human and bots. We found that texts of human are significantly different from those of bots on the other side than the same side in both language communities. This reveals that retweet actions by human correlated with those by bots on the same side to a much higher degree than by the bots of the other side, and partially confirmed our assumption that human tend to follow bots in the social media war on X.

The battlefield of social media war between Ukraine and Russia is dominated by social bots on X. Although the volumes of both #retweet and # users of English language community are significantly larger than in Japanese, we are hesitating to say English are more influential than Japanese. We find that the cascade size of pro-Ukraine bots and pro-Russia bots of Japanese language are higher than their English counterparts, respectively, the # retweeting times show a similar tendency of such difference. We analysed the bot engagement and found that human exhibited a greater inclination to interact with Japanese bots compared to English bots in every efficacy metric of engagement. These results confirm that bots in the Japanese language community played a non-negligible role during the early stage of the Russo-Ukrainian social media war.

Reddit bots correlated human, and pro-Ukraine users are more popular

Similar to X, the majority of the users in German (96.6%) and English (95.0%) are pro-Ukraine. However, human of Reddit are dominating in both German and English language communities (**Fig. 7**). 76.5% of users are human in German language community, and the counterpart in English language community is 79.9%.

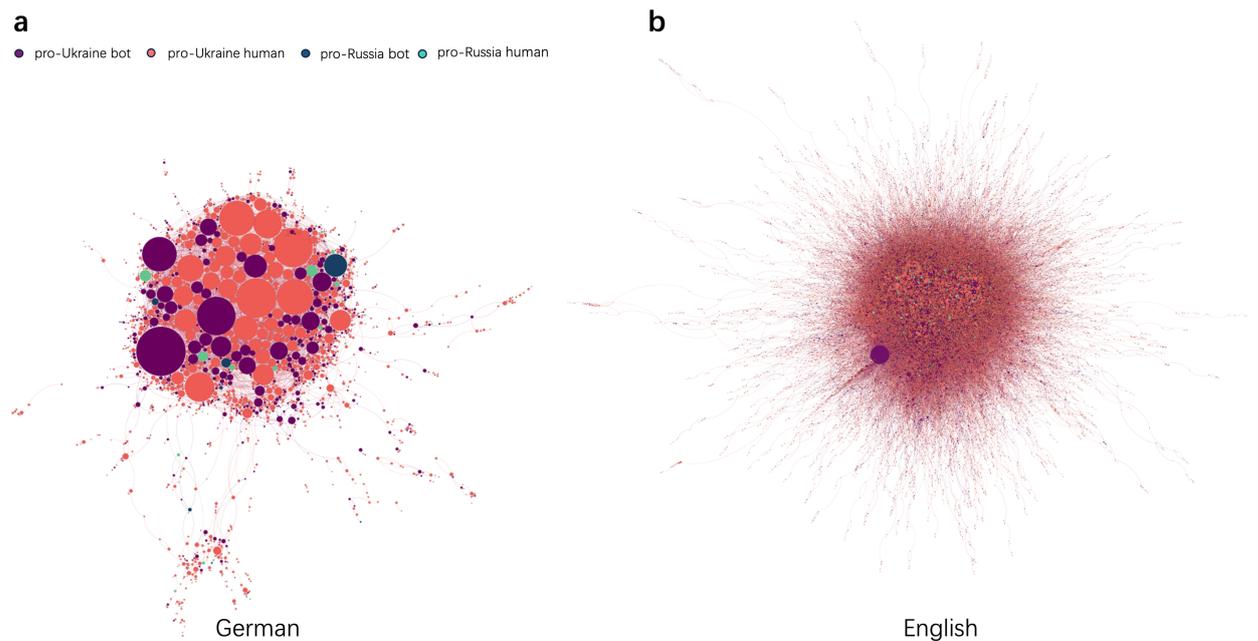

**Fig. 7 | Reddit reply to networks for German (a) and English (b) language community.** Human are dominating in German and English language communities, and the most frequently replied users are bots. For clearer visualization, only labelled pro-Ukraine bots, pro-Ukraine human, pro-Russia bots and pro-Russia human are visualized.

Overall, Reddit bots exceeded human in terms of cascade size (#user counts), but human reached a deeper cascade depth (#reply times). In the English language community, the cascade size of the pro-Russia bot is significantly greater than that of pro-Russia human, and the counterpart of the pro-Ukraine bot is significantly greater than the counterparts of the pro-Ukraine human (**Supplementary Table S8**), indicating bots are more influencing than human on Reddit. For interlanguage analysis, the cascade size of English pro-Ukraine human is significantly larger than the German counterpart. Further investigation showed that the cascade size of the pro-Russia bot is significantly greater than the pro-Russia human in the English language community (*t*-test, $p<0.005$), and the counterpart of pro-Ukraine bots is significantly greater than pro-Ukraine human, as well (*t*-test, $p<0.001$). The statistical analysis suggested that when we looked at the interlanguage difference, the cascade size of English pro-Ukraine human is significantly greater than German counterpart (*t*-test, $p<0.05$).

While considering cascade depth (**Supplementary Table S9**), German pro-Ukraine bot is significantly greater than the English counterpart (*t*-test, $p<0.05$). In addition, German pro-Ukraine human are significantly greater than their English counterparts. We did not identify any statistical significance within the German language community. However, we identified, in the English language community, the cascade depth ranking from largest to lowest: pro-Ukraine human, pro-Ukraine bots, pro-Russia human, and pro-Russia bots. The results suggest that pro-Ukraine influencers are popular on Reddit. Similar to X, users on the same side (Pro-Ukraine / Pro-Russia) exhibit a tendency to parallel each other in terms of #unique users (**Supplementary Fig. S4**). We quantitatively characterised this observation using the Pearson correlation coefficient (**Supplementary Table S10**). The results are consistent with X, indicating that bots exhibit a significantly higher correlation with humans on the same side compared to humans on the other side in both languages.

We then examined the reply to dynamics on Reddit of each user class (**Fig. 8**). We found that bots and human correlated on a high level, except German bots and human. Similar with X, we found that the daily cascade size of English would Granger cause the counterpart of German language community. Through the statistical results, we rejected the null hypothesis and concluded that the English count Granger causes German counterpart, since all of the *p*-values are below the significance level of 0.05, and there is no reverse causality (**Supplementary Table S11**).

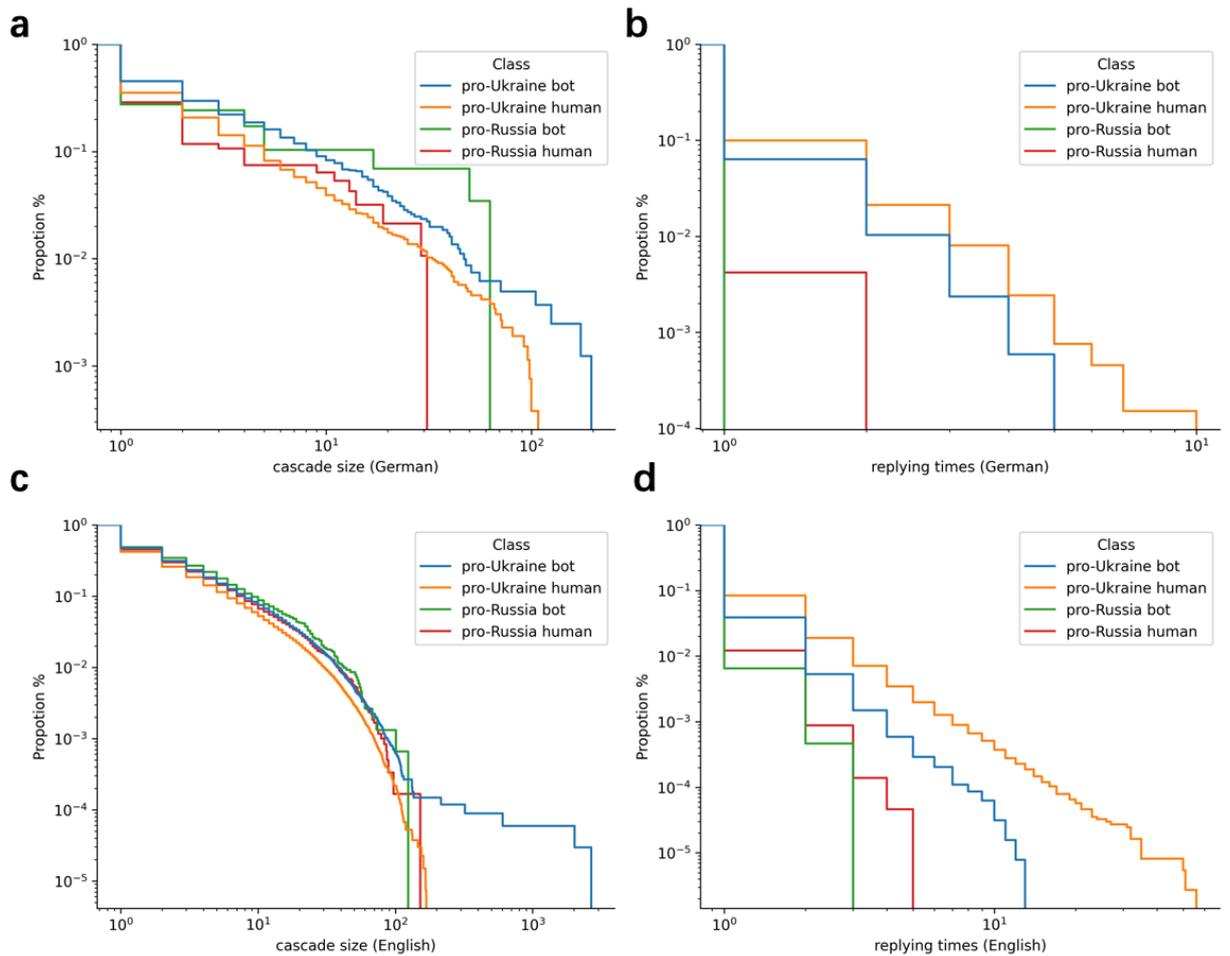

**Fig. 8 | CCDF of the cascade size and the cascade depth of German and English language communities on Reddit.** Panel **a** (cascade size) and panel **b** (cascade depth) are German language community, and the panel **c** and **d** are English counterparts.

We described the reply rate from human to bots in each language community (**Supplementary Table S12**). The rate of German pro-Ukraine human is significantly greater than the counterpart in the English language community, suggesting German humans are more likely to engage with bots. However, we did not identify any pro-Russia human replying to pro-Russia bots within the German language community, compared to English counterparts.

## The bots are playing as hubs on Reddit

Similar to X, co-replying network can be applied in evaluating the bot activity during the early propagation on SNSs. Bots and human synergistically replied on Reddit, and the bot seems to play as a replying hub in the network in German (**Fig. 9**) and English (**Fig.10**) language communities. The co-reply rate between bots and human is second largest to that between human (**Supplementary Table S13**, **S14**). The bot-human co-reply rate is higher within the German-language community compared to the English-language counterpart, across the five time-windows. Although human was increasing and dominating as the time windows moved forward, the role of bots cannot be ignored.

Bots of German and English language communities were identified to become the information hubs in the network for spreading viral messages. The centrality and clustering coefficient of German users are generally higher than English counterparts (*t*-test, $p<0.0001$), indicating German users and their direct neighbours are more connected.

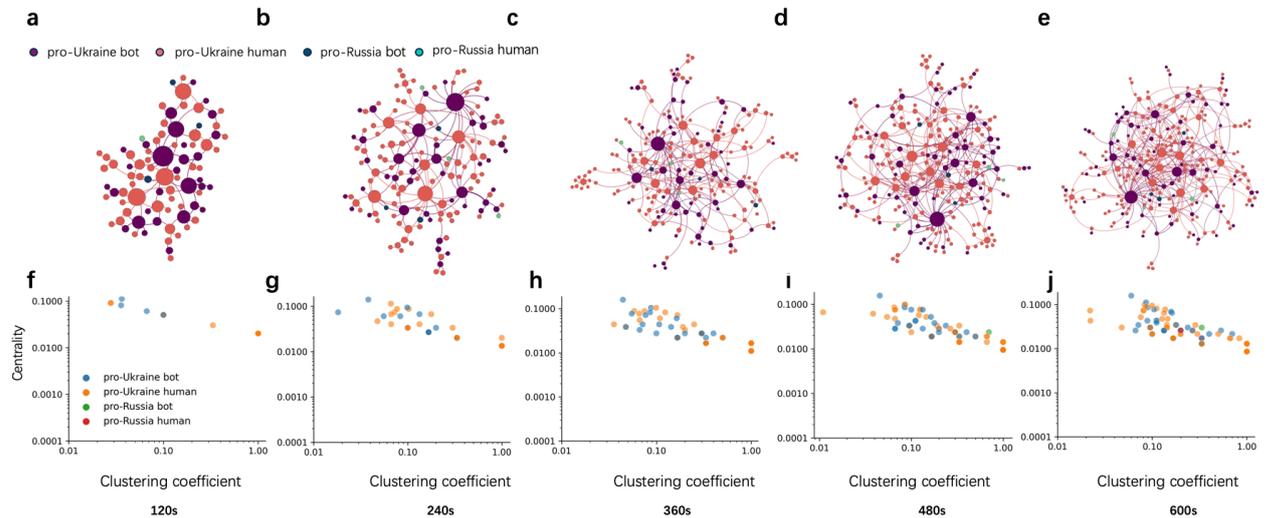

**Fig. 9 | Co-reply network of German language community on Reddit.** The time windows range from 120s to 600s with an interval of 120s. The maximum Indegree users are bots across the time windows. Bots and humans are well-connected across the time windows (**a-e**), and the overall degree of centrality of nodes is higher (**f-j**). Note that the size of a node is scaled by its degree. The visualization scalability is validated only in each time window, respectively.

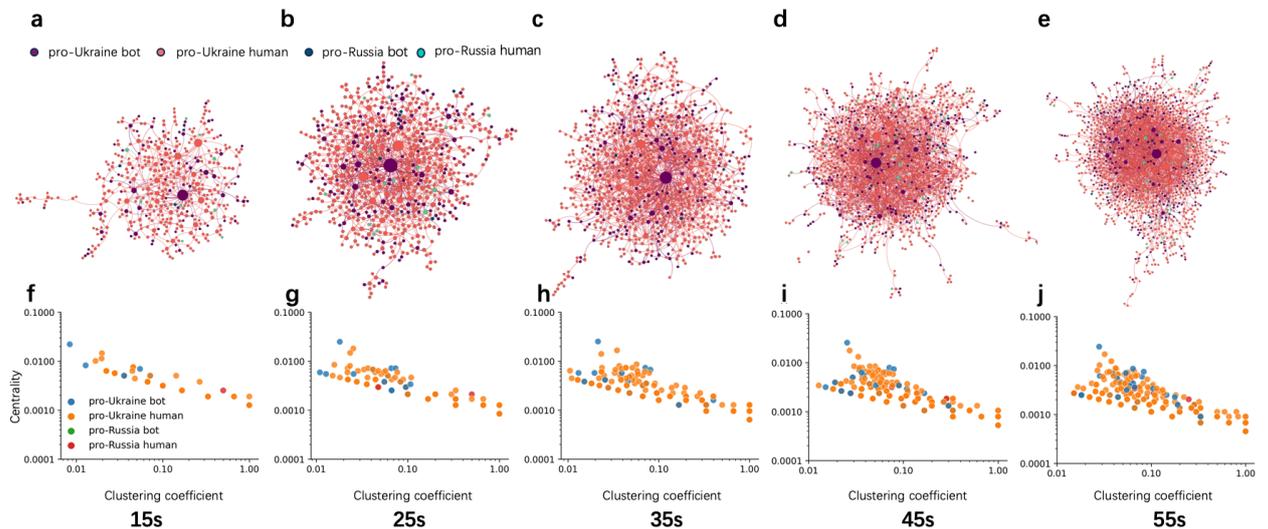

**Fig. 10 | Co-reply network of English language community on Reddit.** The time windows range from 15s to 55s with an interval of 10s. Most of the co-replying users are human, as well. Bots and humans are well-connected across the time windows (**a-e**), and the overall degree of centrality and clustering coefficient of nodes are lower than in German (**f-j**). The bot is the most connected node in each time window, but its immediate neighbours are not well-connected to each other. Note that the size of a node

is scaled by its degree. The visualization scalability is validated only in each time window, respectively.

Bots are toxic spreaders on Reddit

The overall English language toxicity is significantly greater than German on a class pairwise level (**Supplementary Table S15**). However, in German language community, the pro-Ukraine bots are significantly greater than the pro-Russia bots, and both pro-Russia bots and pro-Ukraine bots are significantly greater than pro-Ukraine human. Different from X, German bots were not found to be replied to by human (**Supplementary Table S10**). In addition, the toxicity of German pro-Ukraine human is significantly different from English pro-Ukraine bots than German pro-Ukraine bots, specifically, the pro-Ukraine bots are significantly greater than pro-Ukraine human. Regarding the English language community, 72% of human replied to bots (**Supplementary Table S12**), and the bots are the largest-degree users across the time window (**Fig. 11**). These findings suggested that the bots might be strong toxic content spreaders on Reddit, although they are the minority.

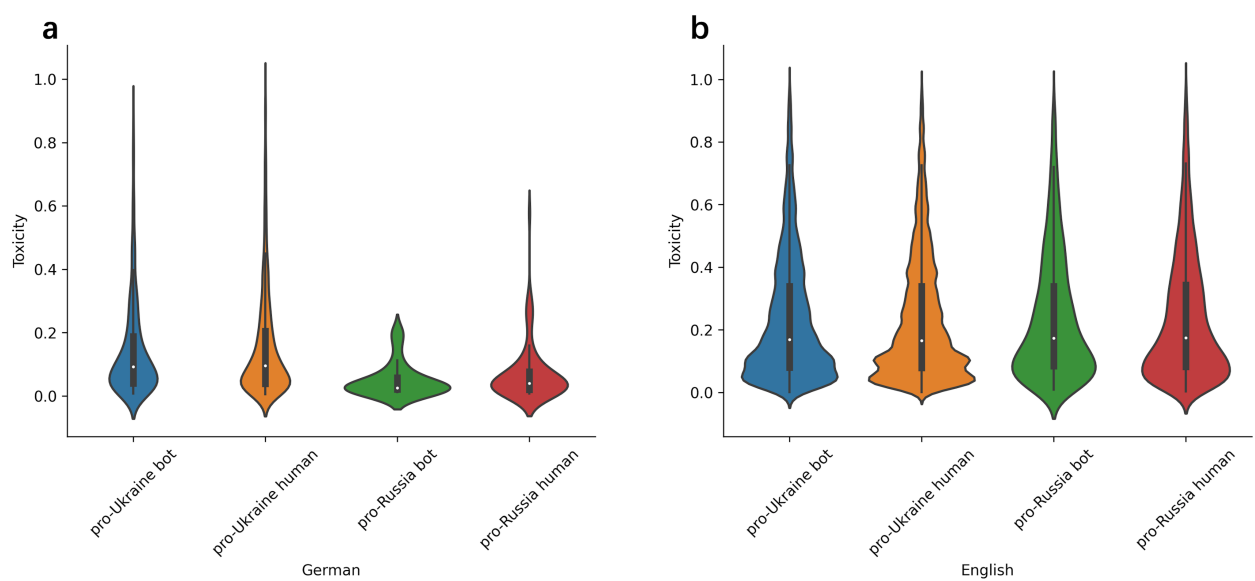

**Fig. 11 | Violin plots represent language toxicity of pro-Ukraine bots, pro-Ukraine human, pro-Russia bots and pro-Russia human of German (a) and English (b) language communities.**

Discussion

We proposed the framework to investigate the bots' pluripotency in the early stage of the contemporary social media war between two major opponents. To measure the bot pluripotency, we classified users into pro-Ukraine bots and pro-Ukraine human, pro-Russia bots and pro-Russia human, and then employed the user engagement, retweet/reply network and language toxicity to investigate the influence, social networking, and linguistic features of bots on various SNSs across multi-languages. The framework shows the efficiency in describing the role of bots during the modern social media war between two major opponents in the world. Our proposed framework provides a rapid snapshot of the evolution of the bot ecosystem on SNSs. This framework has the capability to digitally profile bots during the early stages of the contemporary social media war between Russia and Ukraine.

The majority of users are bots on X, while their counterparts are human on Reddit, suggesting the heterogeneous nature of different SNSs. However, in both X and Reddit, the English language community takes a leading role in Granger causality, with bots playing a central role across various time windows. These findings suggest that the social media war between Ukraine and Russia has attracted such significant global attention that homogeneity governs the trend of bot pluripotency, in terms of influencing, networking and linguistics.

The cascade size and the cascade depth are balanced metrics between importance and continence. The structure of online diffusion cascades has received much attention. The research compared the structure of cascades of different content types[60,61], but rare attempts have been made to integrate them under a framework to digitally profile social bots' delivery mechanisms of different information diffusion types of contemporary social media war. Here, we showed that the cascade size and the cascade depth provide a useful and continent approach to analyse the pluripotency of bots. Our framework is able to identify specific features on different SNSs. For instance, bots were dominating, and Japanese bots were much more influential on X. By contrast, although human prevailed on Reddit, English bots were much more influential than human on Reddit. Moreover, our framework is able to perform statistical causality between languages.

The network visualization performs straighthood snapshots for understanding the dynamics of user engagement. We analysed the structure of the information diffusion network including the retweet and reply network and co-strategy network, such as co-retweet and co-reply network. The four classes of users were clearly identified in the network, and the interaction dynamics in the short time window helped to understand the bot-human engagement on both X and Reddit. Co-retweet network of X showed that bots were co-working fast to become a single retweeting bots network structure, suggesting a balance between strong local cohesion and the presence of influencers that connect smaller communities, potentially facilitating the flow of pro-Ukraine information across the whole network. Compared with the co-retweet network structure, the relatively smaller and isolated Japanese co-reply network suggests that co-reply is not the major pipeline for information flow. Due to the fundamental difference in social networking, the network analysis shifts from the retweet network on X towards reply network on Reddit. The co-reply network of Reddit is rather a "human network", although the bot was still influencing human. Both German and English co-reply networks showed compact clusters, indicating that Reddit human and bots shared many common comments and links and that the replying interaction between human and bots in the German language community is much higher than in English counterparts. Our framework suggests that minor-language communities of X and Reddit are more gated, and the echo chamber might be easier to occur.

Furthermore, English co-retweet network formation is a transition process on X. During this process, the pro-Ukraine bots dominated the network structure with stronger local clustering shifts towards a more centralized and potentially polarized network structure, where certain pro-Ukraine bots are gaining influence, but the overall community structure is weakening. This suggests that the complex network of pro-Ukraine bots might be decentralized as English language community is more than an open community, voices from different sides would dilute the information flow of pro-Ukraine bots. Although the English co-reply network is occupied by pro-Ukraine bots, clusters of in-between bots replies and in-between human' replies are separated, suggesting that an echo chamber might occur in each cluster.

The time-dependent activity, human-bot engagement, and co-strategy behaviours have implied the pluripotency of social bots. Research has demonstrated that social bots are able to generate semantic content depending on the polarized stance of their human influencer targets[59]. Under the context of fierce scenario of social media war, we demonstrated that social bots are further able to produce more toxic texts than human on both X and Reddit. Previous research found that emotions can influence the language used in online discussions, potentially leading to more toxic language[6] and that exposure to toxic language in comments increases the toxicity of subsequent comments[63]. It is possible that the pluripotency helps bots to be differentiated into different language communities on X and Reddit. The level of toxic content helped keep the echo chamber of the social media war across multi-languages. For example, the toxicity of Japanese and German is quite different from English counterparts on X and Reddit, respectively. The low level of openness of the minor language community would help prevent the toxic context from English language community. The unlawful Bucha civilian killings during Russo-Ukrainian war provide people a serious opportunity to look into the synergetic mechanism of pluripotency of social bots. During the early stage of the social media war, bots had people with different camps followed and replied, potentially increasing the confrontation between pro-Ukraine and pro-Russia sides.

Considering the aforementioned attributes of the pluripotency of bots, which enable bots to participate in contemporary social media warfare across different language communities, two significant implications emerge. First, bots have been differentiated to be acting faster, more complexed, and have kept the capability of taking multiple tasks to deeply affect human. SNSs, influencers, researchers, users, government, and other related stakeholders should alert the evolution of bot pluripotency. Large, resourceful organizations are encouraged to adopt an open stance towards research endeavours, establishing collaboration in the advancement of countermeasures against bot pluripotency. Second, human of minority language communities should shape information diffusion in the network of bots. Pro-Ukraine human of Japanese language community on X and those in German and English language communities on Reddit are expected to prepare for the deeper, faster, and more toxic wave of information flow attributed to the bot pluripotency. The bots' pluripotency has evolved with the advancement of large language models. The most recent research identified the emergence of AI-enabled social bots that utilize large language models such as ChatGPT to generate human-like content on X[64]. The study also found that the accounts in the network of bots form a dense cluster and engage with each other through replies and retweets, promoting suspicious websites and spreading harmful comments. The AI capability would further accelerate the pluripotency differentiation of bots being more human-like. Those findings, including ours, underscore the importance of maintaining a high level of vigilance concerning the pluripotency of social bots.

Armed with countermeasures rooted in the characteristics of bots' pluripotency, like the ones we have proposed, we can enhance our defences against targeted attacks by pluripotent bots in the early stage of contemporary social media warfare across multi-language communities.

## Methods

### Framework for bots' pluripotency evaluation

Pluripotency is originated in embryonic stem cell research, referring to the ability of differentiating into various types of cells in human bodies, under appropriate environmental conditions. We extend this concept beyond biomedical research, as bots can be characterised as pluripotent when they signify their ability to attract audiences, disseminate misinformation, and produce human-like language[64].

To assess bot pluripotency, we propose a quantitative empirical framework encompassing the analysis of reposts, network dynamics, and linguistic patterns. This framework has the capability to generate clear and insightful visual representations as well as statistical assessments. The framework contains the measurement of human-bot engagement[65], co-strategy analysis (co-retweet and co-reply[66]), and language toxicity analysis facilitated by Google Perspective API. Human-bot engagement evaluates the popularity of retweets and the frequency of retweeting, while the co-strategy network provides insights into the flow of information among users. Additionally, language toxicity analysis offers valuable insights into linguistic features and potential relationships between bots and human.

These three modules have the capacity to establish interrelated explanations about the social media war, ensuring that each result is as comprehensible and explanatory as possible.

### Data collection

To validate the proposed framework, we expected it would perform on various social media platforms with different social mechanisms. In this study, X and Reddit were selected, as they share different social network service features. X is well known for its "retweet", while "submission" and "comment" are two major features of Reddit. We do not carry out a typical "big data" analysis, our focus is on the early stage of the social media war and thus, we concentrate on one of the most controversial topics – the unlawful civilian killings in Bucha. We employ a one-month time window to gather data that is as closely related as possible to this topic. Traditionally, English language community is believed to be a major battlefield of the war on social media, and the battlefield for minor language is seldom reported. Here, to evaluate that our proposed framework works in complex and diverse language contexts, we carried out experiments in two language communities in each platform, i.e., English and Japanese for X, and English and German for Reddit. Japanese was chosen due to the fact that it is a geopolitically neighbour of Russia, while the country is a close ally of U.S. Studying the Japanese language community can help to understand whether they are affected by English or maintain an independent stance on the topic. German is selected as Germany has a long history of cooperation with Russia, especially in energy. Meanwhile, Germany is a member of NATO. We expected German community on Reddit would hold a balanced position on the topic.

(1) X

We used Twarc, a command line tool for collecting X data via the X API, to collect Bucha-related English tweets from March 28 to April 30, 2022, and their counterparts in Japanese. The query terms for this search include Bucha-related keywords: ("Bucha" AND "Russian") OR ("Bucha" AND "Russia") OR ("Bucha" AND "Ukraine") OR ("Bucha" AND "Ukrainian"). The resulting volume of English tweets was 2,343,641 of which 2,030,029 (86.7%) were retweets (RTs). The number of unique users was 554,303. For Japanese tweets, we harvested 418,375 tweets, of which 370,996 (88.7%) were RTs, and the number of unique users was 89,522. Because the focus of the study is on the tweets about unlawful killings in Bucha in the context of the Russo-Ukrainian War, and thus, we did not intentionally use potentially political-biased words, such as "massacre", "invasion" and "neo-Nazis" in the search query during the data collection to keep the viewpoint of this study as neutral as possible.

(2) Reddit

We used Reddit data archived by Pushshift (https://the-eye.eu/redarcs/). By querying "Bucha OR Russia OR Ukraine" and German counterparts, we collected 1,047,291 and 13,188 for English and German comments, respectively, and we finally labelled 176, 539 English and 3, 570 German users.

## Translation

For pro-Ukraine and pro-Russia stance detection and language toxicity measurement with Google Perspective API, we used deep-translator (https://github.com/nidhaloff/deep-translator) to translate contents in Japanese and German on X and Reddit into English, respectively.

## Classification of users

(1) X user Classification

Since the two ostensible sides of the war are Russia and Ukraine, we expected to identify a characteristic RT network where pro-Russia and pro-Ukraine users are segregated. We produced an RT network using the X data and applied the $k$-core decomposition ($k$=3)[67] to identify users of each side to keep users with only strong connections to the primary discussions. In the network, each node represented a user and the directed edge between two nodes represented a retweet. The modularity-based community detection algorithm, Louvain method[68] was then used to identify the separated clusters.

To verify the results of community detection, a pro-Ukraine/pro-Russia user verification process was conducted by two independent coders. They evaluated 120 randomly selected accounts, with 60 (30 English, 30 Japanese users) labelled as pro-Russia and 60 labelled as pro-Ukraine, by reading their tweets and profiles, and then classified them as pro-Russia or pro-Ukraine users. The resulting kappa=0.70 indicates a significant match between the two coders, indicating that the user separation by the Louvain modularity detection method is sufficiently reliable. Note that according to Landis and Koch[69], Cohen's kappa value is interpreted as follows: 0.0~0.2 for slight agreement; 0.2~0.4 for fair agreement; 0.4~0.6 for moderate agreement; 0.6~0.8 for substantial agreement; and 0.8~1.0 for near perfect agreement.

Next, we used the Botometer API v4 to classify users into human and bots. The Botometer is a well-known tool for identifying social bots based on a supervised machine learning framework. The Botometer has been applied in a series of studies to quantify the online behaviours of bots[70,71]. The Botometer model is trained with 1200 features, covering six categories including the account's metadata, retweet and mention networks, temporal features, content information, and sentiment[72].

For users in English language community, we used "complete automation probability" (CAP) for each user that ranges within [0,1]. This approach takes into consideration the specific linguistic features of the English language. Whereas, for users in Japanese language community, we used the language-independent universal Botometer score instead of CAP, as they are the scores used by social scientists in their studies and are independent from the language that the accounts tweet in. For both type of scores, the higher the value, the higher the probability that the user is a bot. For both types of scores, the higher the value, the higher the probability that the user is a bot.

In this study, we set Botometer score = 0.7 as the threshold for human / bot classification, which means if a user's Botometer score is larger than 0.7, the user is considered to be a bot; the user is human, otherwise. Then, indegrees (the number of retweeted posts by different users, a metric of engagement) as a function of the Botometer score was measured.

Based on the identification of pro-Russia and pro-Ukraine clusters, and bot classification, four categories of users are determined, including, pro-Russia bots, pro-Russia human, pro-Ukraine bots and pro-Ukraine human. For visibility, we illustrated the RT network by including only the four types of accounts using the software Gephi[73], with the graph layout algorithm ForceAtlas2[74].

(2) Reddit user classification
Reddit-spam-bot-detector (https://github.com/creme332/reddit-spam-bot-detector) was used to for human/bots classification. Six features including account age, karma, verified account, Reddit employee, variance in time interval between posts/comments and variance between posts'/comments' content are considered by the algorithm. The algorithm produces True (bots) and False (human) labels for each account, as long as the account is not suspended or closed.

Being different from X platform, Reddit is rather a reply-action social network. Therefore, retweet (repost) network approach is not appropriate for pro-Russia and pro-Ukraine community classification. Here, we used contextual information of Reddit

"body" (text) object to label each user. Two independent coders were invited to label each comment text with "pro-Ukraine", "pro-Russia" and "Unknown" from 500 randomly selected comments. The kappa agreement for "pro-Russia" and "pro-Ukraine" is 0.73. We then used Autolabel (https://github.com/refuel-ai/autolabel) to classify each user. The comment (reply) texts were aggregated for each user, representing the user's stance. We then created a labelling agent with Autolabel. The agent used Google T5-small[75] to carry out labelling. To enhance the agent performance, we used chain-of-thought prompting technique[76], in which 10 "pro-Ukraine" and 10 "pro-Russia" labelled by the coders were randomly selected as learning examples for the label agent. The labelling agent was running on two local NVIDIA RTX 3090 GPUs with 48GB of VRAM.

## Measurement of influencing

On X platform, a retweet can be both a productive, communicative tool and a selfish act of an attention seeker[65]. Retweeting is a simple and powerful action for amplifying information from the influencer who creates the original message of an information cascade. In this study, we assume that a Twitter message cascade is originated by a unique influencer posting a unique URL. We aggregated all unique influencers and corresponding URLs, and computed user-level influence as the logarithm of the size of all cascades of four classes of users, including "pro-Ukraine human," "pro-Ukraine bots," "pro-Russia human," and "pro-Russia bots."

The cascade size in X scenario is defined as the number of retweeted times of a unique URL created by a unique influencer. For instance, if 3 URLs, "A," "B," and "C" are tweeted by a single influencer, then we have three cascades, and the cascade size is 3. Conversely, in the same example, if each URL is further retweeted by 10 times, then we still have three cascades, but cascade size is 30.

The cascade depth is defined as the retweeting times, as the "reposting" times[77], is also an important metric for measuring X cascade influence. Obviously, the more a message is retweeted, the more likely that message is to be acknowledged. However, bots can work synergistically to retweet a message on purpose. Bots are able to retweet articles within seconds of their first being posted, leading to the articles going viral[70]. Here, we simply use the retweeting times of a unique influencer to measure the popularity of the influencer to the four classes of users, i.e., the cascade depth is able to measure the number of unique users retweeting an influencer over time. A deeper cascade in X suggests that the original tweet or topic of an influencer has reached a broader audience. This indicates a higher level of exposure and visibility for the content and can also indicate greater engagement with the influencer.

On the Reddit platform, the cascade size of a user simply refers to the aggregated appearance of each user class, i.e., we aggregated all unique users and counted their appearance on the user class level as the logarithm of the size. The cascade depth on Reddit is defined as the number of replied times of a unique "parent_id" of a subreddit, i.e., the times of replying to the link or comment in the id. A greater cascade size indicates the user class is replying more frequently, and a deeper cascade depth would suggest the user class is more attracted by the other users.

## Granger causality test

Granger causality test[78] is a statistical test used to determine whether one time series can be used to predict another. The test is based on the concept that if a time series X Granger causes another time series Y, then the past values of X provide useful information for predicting the future values of Y. The Granger causality test has proven to be a valuable tool in understanding causal relationships between time series data and has been applied in large-scale functional magnetic resonance imaging data analsysis[79], earth system sciences[80] and groundwater patterns[81]. Please note that Granger causality may not necessarily establish a direct cause-and-effect relationship. In this study, we used statsmodels (https://github.com/statsmodels/statsmodels) to carry out the Granger causality test to measure the potential causal relation of cascade size and depth between English and Japanese / German language communities. The max lag is set to five, as default.

## Measurement of user engagement

In order to measure the efficiency of human – social bot engagement of the pro-Russia and pro-Ukraine clusters, we adopt three metrics for X from the research conducted by Luceri and his colleague[22]. The primary objective of these metrics is to assess the degree to which individuals depend on and engage with content that is either generated or facilitated by social bots.

The metrics include:

1. Retweet pervasiveness (RTP) measures the extent of penetration of bot-generated content into human. $RTP$ is defined as:

$$RTP = \frac{\#\ of\ human\ retweets\ from\ bot\ tweets}{\#\ of\ humans\ retweets} \quad (1)$$

2. Reply rate (RR) quantifies the number of replies given by human to social bots on X. $RR$ is defined as:

$$RR = \frac{\#\ of\ human\ replies\ from\ bot\ tweets}{\#\ of\ human\ replies} \quad (2)$$

3. Human to Bot Rate (H2BR) measures the human interaction with bots over all human activity in the network. $H2BR$ is defined as:

$$H2BR = \frac{\#\ of\ human\ interaction\ with\ bots}{\#\ of\ human\ interaction} \quad (3)$$

For Reddit platform, we only used a simplified the metric of $RR_X$, to measure the engagement of human and bots. $RR_{reddit}$ is defined as:

$$RR_{reddit} = \frac{\#\ of\ human\ replies\ from\ bot\ comments}{\#\ of\ human\ replies} \quad (4)$$

Please note that one user may retweet/reply to multiple categories of bots.

## Networks analysis

For X platform, co-occurred patterns can help us visualize conspiracy theory meta umbrella[82], and help generate network user embeddings[83], which further inspired us to use co-strategies including co-retweet and co-reply to understand the nature of human-bot interactions. A co-retweet or co-reply network refers to a network where individual users are connected based on their co-occurrence in retweets and replies activities. These networks can be studied using various methods and measures. For example, one study investigated the social network of co-occurrence in a news corpus, representing people as vertices and connecting them if they co-occur in the same article[84]. In this study, we constructed co-retweet network and co-reply network by using "Coordination Network Toolkit"[66]. The default time window of the tool is 60 seconds for all network types. To examine the dynamics of co-retweet network, we used 15~55 seconds time window with a 10-second interval. For the visualization of dynamics of co-reply networks, we set 1 day, 7 days and 14 days as time windows. To measure the network dynamics of Reddit, we constructed the German co-reply networks by using 120~600 seconds time window with a 120-second interval and generated the English co-reply networks by using 15~55 seconds time window with a 10-second interval. Please note that, due to the Reddit data limitation, we only considered the scenario of co-replying, in which a user replies to a subreddit created by another user. This limitation might explain why German human was not identified to reply to bots in **Supplementary Table S12**.

To quantitatively measure the network, we analysed clustering coefficient against centrality (degree) of four user classes. The clustering coefficient and centrality are two important measures for analysing social networks. The clustering coefficient measures the extent to which users in a network tend to cluster together, indicating the presence of communities or tightly connected groups. On the other hand, degree centrality measures the number of connections or interactions that a user has, indicating the user's importance or influence within the network. In the context of X networks, several studies have explored the relationship between these two measures. One study found a low global clustering coefficient average and a power law distribution for the degree centrality[85]. Another study observed that power laws apply for the influencer degree centrality in topical networks on X[86]. Here we used NetworkX to compute the clustering coefficient and centrality of each user.

## Measurement of language toxicity

In order to measure the language association between bots and humans, we quantified users' toxicity. A user's toxicity is calculated by measuring the language toxicity of aggregated text (tweets of X, and comments of Reddit) of a user. The language toxicities were measured by Google's Perspective API.

# References

1. Minsky, M. L. Emotion. In *The Society of Mind* 163 (Simon & Schuster, New York, 1988).
2. Shao, C. *et al.* The spread of low-credibility content by social bots. *Nat Commun* **9**, (2018).
3. Stella, M., Ferrara, E. & Domenico, M. De. Bots increase exposure to negative and inflammatory content in online social systems. *Proc Natl Acad Sci U S A* **115**, 12435–12440 (2018).
4. Santia, G. C., Mujib, M. I. & Williams, J. R. Detecting social bots on facebook in an information veracity context. In *International Conference on Web and Social Media* (2019).
5. Seering, J., Flores, J. P., Savage, S. & Hammer, J. The social roles of bots: Evaluating impact of bots on discussions in online communities. *Proc. ACM Hum.-Comput. Interact.* **2**, (2018).
6. Tsvetkova, M., García-Gavilanes, R., Floridi, L. & Yasseri, T. Even good bots fight: The case of Wikipedia. *PLoS One* **12**, e0171774 (2017).
7. Caldarelli, G., De Nicola, R., Del Vigna, F., Petrocchi, M. & Saracco, F. The role of bot squads in the political propaganda on twitter. *Commun Phys* **3**, 81 (2020).
8. Takahashi, K. & Yamanaka, S. Induction of pluripotent stem cells from mouse embryonic and adult fibroblast cultures by defined factors. *Cell* **126**, 663–676 (2006).
9. Caldarelli, G., De Nicola, R., Del Vigna, F., Petrocchi, M. & Saracco, F. The role of bot squads in the political propaganda on Twitter. *Commun Phys* **3**, (2020).
10. Bakshy, E., Messing, S. & Adamic, L. A. Exposure to ideologically diverse news and opinion on Facebook. *Science* **348**, 1130–1132 (2015).
11. Ferrara, E., Varol, O., Davis, C., Menczer, F. & Flammini, A. The rise of social bots. *Commun. ACM* **59**, 96–104 (2016).
12. Lazer, D. M. J. *et al.* The science of fake news: Addressing fake news requires a multidisciplinary effort. *Science* **359**, 1094–1096 (2018).
13. Boshmaf, Y., Muslukhov, I., Beznosov, K. & Ripeanu, M. The socialbot network: when bots socialize for fame and money. In *Proceedings of the 27th Annual Computer Security Applications Conference* 93–102 (Association for Computing Machinery, New York, NY, USA, 2011). doi:10.1145/2076732.2076746.
14. Ji, Y., He, Y., Jiang, X., Cao, J. & Li, Q. Combating the evasion mechanisms of social bots. *Comput Secur* **58**, 230–249 (2016).
15. Hasal, M. *et al.* Chatbots: Security, privacy, data protection, and social aspects. *Concurr Comput* **33**, e6426 (2021).
16. Kayes, I. & Iamnitchi, A. Privacy and security In online social networks: A survey. *Online Soc Netw Media* **3–4**, 1–21 (2017).
17. Duan, Z. *et al.* Algorithmic agents in the hybrid media system: Social bots, selective amplification, and partisan news about covid-19. *Hum Commun Res* **48**, 516–542 (2022).
18. Marlow, T., Miller, S. & Roberts, J. T. Bots and online climate discourses: Twitter discourse on president trump's announcement of u.s. withdrawal from the paris agreement. *Clim. Policy* **21**, 765–777 (2021).

19. Keller, F. B., Schoch, D., Stier, S. & Yang, J. Political astroturfing on twitter: How to coordinate a disinformation campaign. *Polit Commun* **37**, 256–280 (2020).
20. Pescetelli, N., Barkoczi, D. & Cebrian, M. Bots influence opinion dynamics without direct human-bot interaction: the mediating role of recommender systems. *Appl Netw Sci* **7**, 46 (2022).
21. Luo, H., Meng, X., Zhao, Y. & Cai, M. Rise of social bots: The impact of social bots on public opinion dynamics in public health emergencies from an information ecology perspective. *Telemat. Inform.* **85**, 102051 (2023).
22. Luceri, L., Deb, A., Badawy, A. & Ferrara, E. Red bots do it better: Comparative analysis of social bot partisan behavior. In *Companion Proceedings of The 2019 World Wide Web Conference* 1007–1012 (Association for Computing Machinery, New York, NY, USA, 2019). doi:10.1145/3308560.3316735.
23. Cheng, C., Luo, Y. & Yu, C. Dynamic mechanism of social bots interfering with public opinion in network. *Physica A: Statistical Mechanics and its Applications* **551**, 124163 (2020).
24. Aldayel, A. & Magdy, W. Characterizing the role of bots' in polarized stance on social media. *Soc Netw Anal Min* **12**, 30 (2022).
25. Zhou, M., Zhang, D., Wang, Y., Geng, Y.-A. & Tang, J. Detecting social bot on the fly using contrastive learning. In *Proceedings of the 32nd ACM International Conference on Information and Knowledge Management* 4995–5001 (Association for Computing Machinery, New York, NY, USA, 2023). doi:10.1145/3583780.3615468.
26. Ferrara, E. Social bot detection in the age of chatgpt: Challenges and opportunities. *First Monday* **28**, (2023).
27. Xu, W. & Sasahara, K. Characterizing the roles of bots on twitter during the covid-19 infodemic. *J Comput Soc Sci* **5**, 591–609 (2022).
28. Varol, O., Ferrara, E., Davis, C. A., Menczer, F. & Flammini, A. Online human-bot interactions: Detection, estimation, and characterization. In *International Conference on Web and Social Media* (2017).
29. Reeves, B., Hancock, J. T. & Liu, X. Social robots are like real people: First impressions, attributes, and stereotyping of social robots. *Technology, Mind, and Behavior* **1**, (2020).
30. Wischnewski, M., Ngo, T., Bernemann, R., Jansen, M. & Krämer, N. C. "I agree with you, bot!" How users (dis)engage with social bots on twitter. *New Media Soc* (2022).
31. Schoeffer, J. *et al.* Online platforms and the fair exposure problem under Homophily. In *Proceedings of the AAAI Conference on Artificial Intelligence* **37**, 11899–11908 (2023).
32. Jana, R. K., Maity, S. & Maiti, S. An empirical study of sentiment and behavioural analysis using homophily effect in social network. In *2022 6th International Conference on Intelligent Computing and Control Systems (ICICCS)* 1508–1515 (2022). doi:10.1109/ICICCS53718.2022.9788407.
33. Alkiek, K., Zhang, B. & Jurgens, D. Classification without (Proper) Representation: Political heterogeneity in social media and its implications for classification and behavioral analysis. In *Findings of the Association for*

*Computational Linguistics: ACL 2022* (eds. Muresan, S., Nakov, P. & Villavicencio, A.) 504–522 (Association for Computational Linguistics, Dublin, Ireland, 2022). doi:10.18653/v1/2022.findings-acl.43.
34. Su, J., Kamath, K., Sharma, A., Ugander, J. & Goel, S. An experimental study of structural diversity in social networks. In *Proceedings of the International AAAI Conference on Web and Social Media* **14**, 661–670 (2020).
35. Jaidka, K. Cross-platform- and subgroup-differences in the well-being effects of twitter, instagram, and facebook in the united states. *Sci Rep* **12**, 3271 (2022).
36. Bailey, E. A. The internet, social media and politics. In *Political Participation on Social Media: The Lived Experience of Online Debate. Political Campaigning and Communication.* (ed. Bailey, E. A.) 19–37 (Springer International Publishing, Cham, 2021). doi:10.1007/978-3-030-65221-0_2.
37. Krzyżanowski, M. & Tucker, J. A. Re/constructing politics through social & online media: Discourses, ideologies, and mediated political practices. *Journal of Language and Politics* **17**, 141–154 (2018).
38. Álvarez Bornstein, B. & Montesi, M. Interdisciplinary research and societal impact: analysis of social media. Preprint at http://hdl.handle.net/20.500.12424/3948232 (2020).
39. Chen, W., Pacheco, D., Yang, K. C. & Menczer, F. Neutral bots probe political bias on social media. *Nat Commun* **12**, (2021).
40. Guzmán Rincón, A., Barragán Moreno, S., Rodríguez-Canovas, B., Carrillo Barbosa, R. L. & Africano Franco, D. R. Social networks, disinformation and diplomacy: a dynamic model for a current problem. *Humanit Soc Sci Commun* **10**, 505 (2023).
41. Gallotti, R., Valle, F., Castaldo, N., Sacco, P. & De Domenico, M. Assessing the risks of 'infodemics' in response to COVID-19 epidemics. *Nat Hum Behav* **4**, 1285–1293 (2020).
42. Johnson, N. F. *et al.* The online competition between pro- and anti-vaccination views. *Nature* **582**, 230–233 (2020).
43. Klein, C. *et al.* Attention and counter-framing in the black lives matter movement on Twitter. *Humanit Soc Sci Commun* **9**, 367 (2022).
44. Wang, E. L., Luceri, L., Pierri, F. & Ferrara, E. Identifying and characterizing behavioral classes of Radicalization within the qanon conspiracy on twitter. In *Proceedings of the International AAAI Conference on Web and Social Media* **17**, 890–901 (2023).
45. Macskassy, S. & Michelson, M. Why do people retweet? anti-homophily wins the day! In *Proceedings of the International AAAI Conference on Web and Social Media* **5**, 209–216 (2021).
46. Wang, Z. Collective memory and national identity. In *Memory Politics, Identity and Conflict: Historical Memory as a Variable* (ed. Wang, Z.) 11–25 (Springer International Publishing, Cham, 2017). doi:10.1007/978-3-319-62621-5_2.
47. Hornsey, M. J., Bierwiaczonek, K., Sassenberg, K. & Douglas, K. M. Individual, intergroup and nation-level influences on belief in conspiracy theories. *Nature Reviews Psychology* **2**, 85–97 (2023).

48. Frith, J., Campbell, S. & Komen, L. J. Looking back to look forward: 5g/covid-19 conspiracies and the long history of infrastructural fears. *Mob Media Commun* **11**, 174–192 (2022).
49. Wang, Y., Wang, X., Ran, Y., Michalski, R. & Jia, T. CasSeqGCN: Combining network structure and temporal sequence to predict information cascades. *Expert Syst Appl* **206**, 117693 (2022).
50. Bhowmick, A. K. Temporal pattern of retweet(s) help to maximize information diffusion in twitter. In *Proceedings of the 13th International Conference on Web Search and Data Mining* 913–914 (Association for Computing Machinery, New York, NY, USA, 2020). doi:10.1145/3336191.3372181.
51. Bhowmick, A. K., Gueuning, M., Delvenne, J.-C., Lambiotte, R. & Mitra, B. Temporal Pattern of (Re)tweets Reveal Cascade Migration. In *Proceedings of the 2017 IEEE/ACM International Conference on Advances in Social Networks Analysis and Mining 2017* 483–488 (Association for Computing Machinery, New York, NY, USA, 2017). doi:10.1145/3110025.3110084.
52. Xie, W., Zhu, F., Liu, S. & Wang, K. Modelling cascades over time in microblogs. In *2015 IEEE International Conference on Big Data (Big Data)* 677–686 (2015). doi:10.1109/BigData.2015.7363812.
53. Elsharkawy, S., Hassan, G., Nabhan, T. & Roushdy, M. Towards feature selection for cascade growth prediction on twitter. In *Proceedings of the 10th International Conference on Informatics and Systems* 166–172 (Association for Computing Machinery, New York, NY, USA, 2016). doi:10.1145/2908446.2908463.
54. Haldar, A., Wang, S., Demirci, G. V., Oakley, J. & Ferhatosmanoglu, H. Temporal cascade model for analyzing spread in evolving networks. *ACM Trans. Spatial Algorithms Syst.* **9**, (2023).
55. Shen, F. *et al.* Examining the differences between human and bot social media accounts: A case study of the Russia-Ukraine War. *First Monday* **28**, (2023).
56. Smart, B., Watt, J., Benedetti, S., Mitchell, L. & Roughan, M. #Istandwithputin versus #istandwithukraine: The interaction of bots and humans in discussion of the russia/ukraine war. *Lecture Notes in Computer Science (including subseries Lecture Notes in Artificial Intelligence and Lecture Notes in Bioinformatics)* **13618 LNCS**, 34–53 (2022).
57. Magelinski, T., Ng, L. & Carley, K. A synchronized action framework for detection of coordination on social media. *Journal of Online Trust and Safety* **1**, (2022).
58. Goodwin, M. H., Cekaite, A. & Goodwin, C. Emotion as stance. In *Emotion in Interaction* (eds. Perakyla, A. & Sorjonen, M.-L.) 16–41 (Oxford University Press, 2012). doi:10.1093/acprof:oso/9780199730735.003.0002.
59. Stella, M., Ferrara, E. & De Domenico, M. Bots increase exposure to negative and inflammatory content in online social systems. *Proc Natl Acad Sci U S A* **115**, 12435–12440 (2018).
60. Flamino, J. *et al.* Political polarization of news media and influencers on Twitter in the 2016 and 2020 US presidential elections. *Nat Hum Behav* **7**, 904–916 (2023).

61. Morone, F. & Makse, H. A. Influence maximization in complex networks through optimal percolation. *Nature* **524**, 65–68 (2015).
62. Xia, Y., Zhu, H., Lu, T., Zhang, P. & Gu, N. Exploring antecedents and consequences of toxicity in online discussions: A Case Study on Reddit. *Proc ACM Hum Comput Interact* **4**, 1–23 (2020).
63. Kim, J. W., Guess, A., Nyhan, B. & Reifler, J. The distorting prism of social media: How self-selection and exposure to incivility fuel online comment toxicity. *J.COMMUN* **71**, 922–946 (2021).
64. Yang, K.-C. & Menczer, F. Anatomy of an ai-powered malicious social botnet. *arXiv e-prints* arXiv:2307.16336 Preprint at https://doi.org/10.48550/arXiv.2307.16336 (2023).
65. Boyd, D., Golder, S. & Lotan, G. Tweet, tweet, retweet: Conversational aspects of retweeting on twitter. In *2010 43rd Hawaii International Conference on System Sciences* 1–10 (2010). doi:10.1109/HICSS.2010.412.
66. Graham, T. & Digital Observatory QUT. Coordination network toolkit (Software). (2020) doi:10.25912/RDF_1632782596538.
67. Giatsidis, C., Thilikos, D. M. & Vazirgiannis, M. D-cores: Measuring collaboration of directed graphs based on degeneracy. In *2011 IEEE 11th International Conference on Data Mining* 201–210 (2011). doi:10.1109/ICDM.2011.46.
68. Blondel, V. D., Guillaume, J.-L., Lambiotte, R. & Lefebvre, E. Fast unfolding of communities in large networks. *J. Stat. Mech. Theory Exp.* **2008**, 10008 (2008).
69. Landis, J. R. & Koch, G. G. The measurement of observer agreement for categorical data. *Biometrics* **33**, 159–174 (1977).
70. Shao, C. *et al.* The spread of low-credibility content by social bots. *Nat Commun* **9**, 4787 (2018).
71. Vosoughi, S., Roy, D. & Aral, S. The spread of true and false news online. *Science* **359**, 1146–1151 (2018).
72. Sayyadiharikandeh, M., Varol, O., Yang, K., Flammini, A. & Menczer, F. Detection of novel social bots by ensembles of specialized classifiers. *ArXiv* **abs/2006.0**, (2020).
73. Bastian, M., Heymann, S. & Jacomy, M. Gephi: An open source software for exploring and manipulating networks. In *Proceedings of the International AAAI Conference on Web and Social Media* **3**, 361–362 (2009).
74. Jacomy, M., Venturini, T., Heymann, S. & Bastian, M. ForceAtlas2, a continuous graph layout algorithm for handy network visualization designed for the Gephi Software. *PLoS One* **9**, e98679 (2014).
75. Raffel, C. *et al.* Exploring the limits of transfer learning with a unified text-to-text transformer. *Journal of Machine Learning Research* **21**, 1–67 (2019).
76. Wei, J. *et al.* Chain-of-thought prompting elicits reasoning in large language models. *arXiv e-prints* arXiv:2201.11903 Preprint at https://doi.org/10.48550/arXiv.2201.11903 (2022).
77. Bakshy, E., Hofman, J., Mason, W. & Watts, D. Everyone's an influencer: Quantifying influence on twitter. In *Proceedings of the 4th ACM International*

*Conference on Web Search and Data Mining, WSDM 2011* (2011). doi:10.1145/1935826.1935845.

78. Granger, C. W. J. Investigating causal relations by econometric models and cross-spectral methods. *Econometrica* **37**, 424–438 (1969).

79. Wismüller, A., Dsouza, A. M., Vosoughi, M. A. & Abidin, A. Large-scale nonlinear granger causality for inferring directed dependence from short multivariate time-series data. *Scientific Reports 2021 11:1* **11**, 1–11 (2021).

80. Runge, J. *et al.* Inferring causation from time series in Earth system sciences. *Nat Commun* **10**, 1–13 (2019).

81. Singh, N. K. & Borrok, D. M. A Granger causality analysis of groundwater patterns over a half-century. *Sci Rep* **9**, 1–8 (2019).

82. Xu, W. & Sasahara, K. A network-based approach to qanon user dynamics during COVID-19 infodemic. *APSIPA Trans Signal Inf Process* **11**, (2021).

83. Xu, W. & Sasahara, K. Domain-based user embedding for competing events on social media. *AXiv Preprint* (2023) doi:10.48550/arXiv.2308.14806.

84. Didegah, F. & Thelwall, M. Co-saved, co-tweeted, and co-cited networks. *J Assoc Inf Sci Technol* **69**, 959–973 (2018).

85. Kustudic, M., Xue, B., Zhong, H., Tan, L. & Niu, B. Identifying communication topologies on twitter. *Electronics (Basel)* **10**, (2021).

86. Schuchard, R., Crooks, A., Stefanidis, A. & Croitoru, A. Bots in nets: Empirical comparative analysis of bot evidence. In *Complex Networks and Their Applications VII. COMPLEX NETWORKS 2018. Studies in Computational Intelligence* (eds. Aiello, L. M. et al.) 424–436 (Springer International Publishing, Cham, 2019).